\documentclass[aps,prl,preprint,groupedaddress]{revtex4}
\usepackage{amsmath,amssymb,graphicx}
\begin{document}
\preprint{\parbox[b]{1in}{ \hbox{\tt PNUTP-09/A03} \hbox{\tt SNUTP09-008}}}

\title{Pseudo scalar contributions to light-by-light correction\\ of muon $g-2$ in AdS/QCD}


\author{Deog Ki Hong}
\email[]{dkhong@pusan.ac.kr}
\affiliation{Department of
Physics,   Pusan National University,
             Busan 609-735, Korea}
\author{Doyoun Kim}
\email[]{abistp00@phya.snu.ac.kr}
\affiliation{FPRD and Department of Physics and Astronomy,\\
 Seoul National University,  Seoul 151-747, Korea}


\date{\today}

\begin{abstract}
	We have performed a holographic calculation of the hadronic contributions to the anomalous magnetic moment of the muon, using the gauge/gravity duality.  As a gravity dual model of QCD with three light flavors, we study a ${\rm U}(3)_{L}\times {\rm U}(3)_{R}$ flavor gauge theory  in the five dimensional AdS background with a hard-wall cutoff.  The anomalous (electromagnetic) form factors for the pseudo scalars,  $\pi^0$, $\eta$ and $\eta^\prime$, are obtained from the 5D Chern-Simons term of the gravity dual, which correctly reproduce the asymptotic behavior of the form factor, dictated by QCD. We find the total light-by-light contributions of pseudo scalars to the muon anomalous magnetic moment, $a_\mu^{\textrm{PS}}=10.7\times 10^{-10}$, which is consistent with previous estimates, based on other approaches.
\end{abstract}

\pacs{}
\keywords{AdS/QCD, muon, muon g-2, anomalous magnetic moment}

\maketitle

\subsection{Introduction}
One of most stringent tests of the standard model (SM) is provided by the measurement of the muon anomalous magnetic moment, $a_{\mu}$, whose current precision is better than parts per million (ppm). Recent measurement of the $(g-2)$ value of the  muon~\cite{Bennett:2006fi}, performed at the Brookhaven National Laboratory (BNL), 
\begin{equation}
a_{\mu}=11 659 208.0(5.4)(3.3)\times10^{-10}\,,
\end{equation}
deviates by $2.2-2.7\,\sigma$  above the current SM estimate, based on $e^{+}e^{-}$ hadronic cross sections. An improved muon $(g-2)$ experiment has been proposed to achieve a precision of $0.1$ ppm~\cite{Miller:2007kk}. 
The discrepancy between the SM estimate and the experimental value, if persists, might hint a new physics beyond the standard model.

\begin{figure}[tbh]
	\centering
	\includegraphics[width=0.5\textwidth]{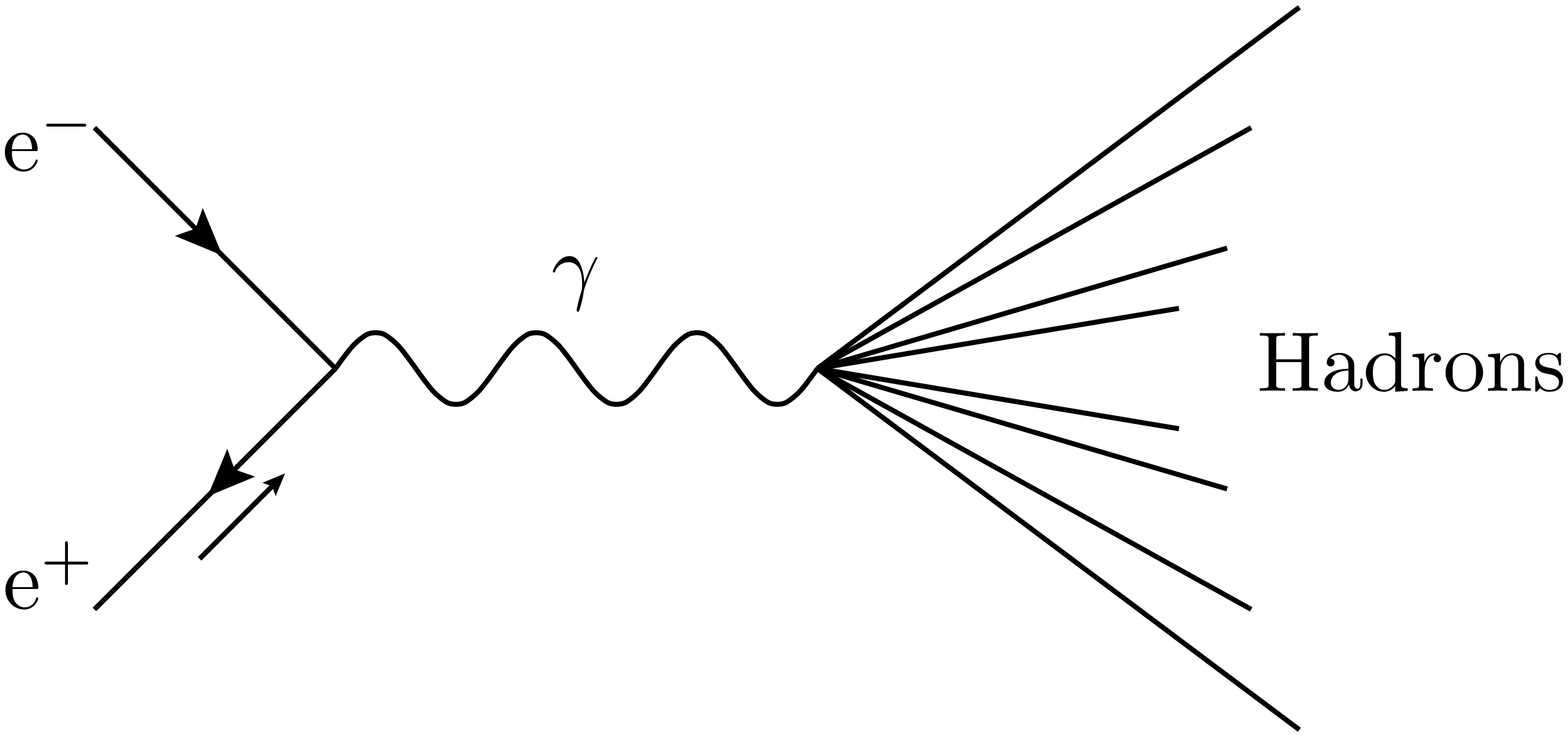}%
	\\
	\includegraphics[width=0.5\textwidth]{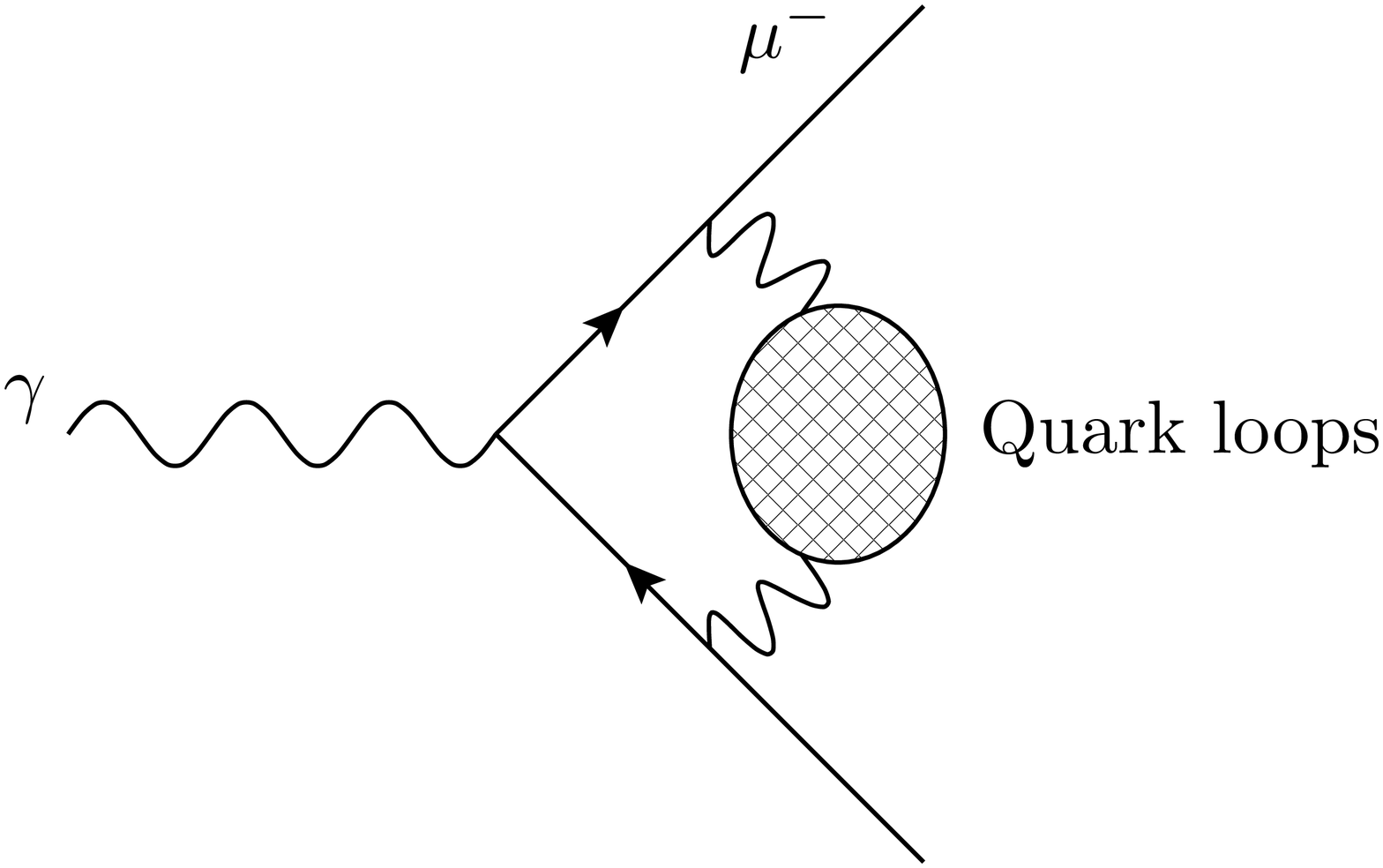}
	\caption{\label{HLO diagram}HLO correction (bottom) essentially involves a dressed propagator of the photon, which is related to  the  process (up) of annihilation into hadrons by unitarity and analyticity.}
\end{figure}

\begin{figure}[tbh]
	\centering
	\includegraphics[width=0.5\textwidth]{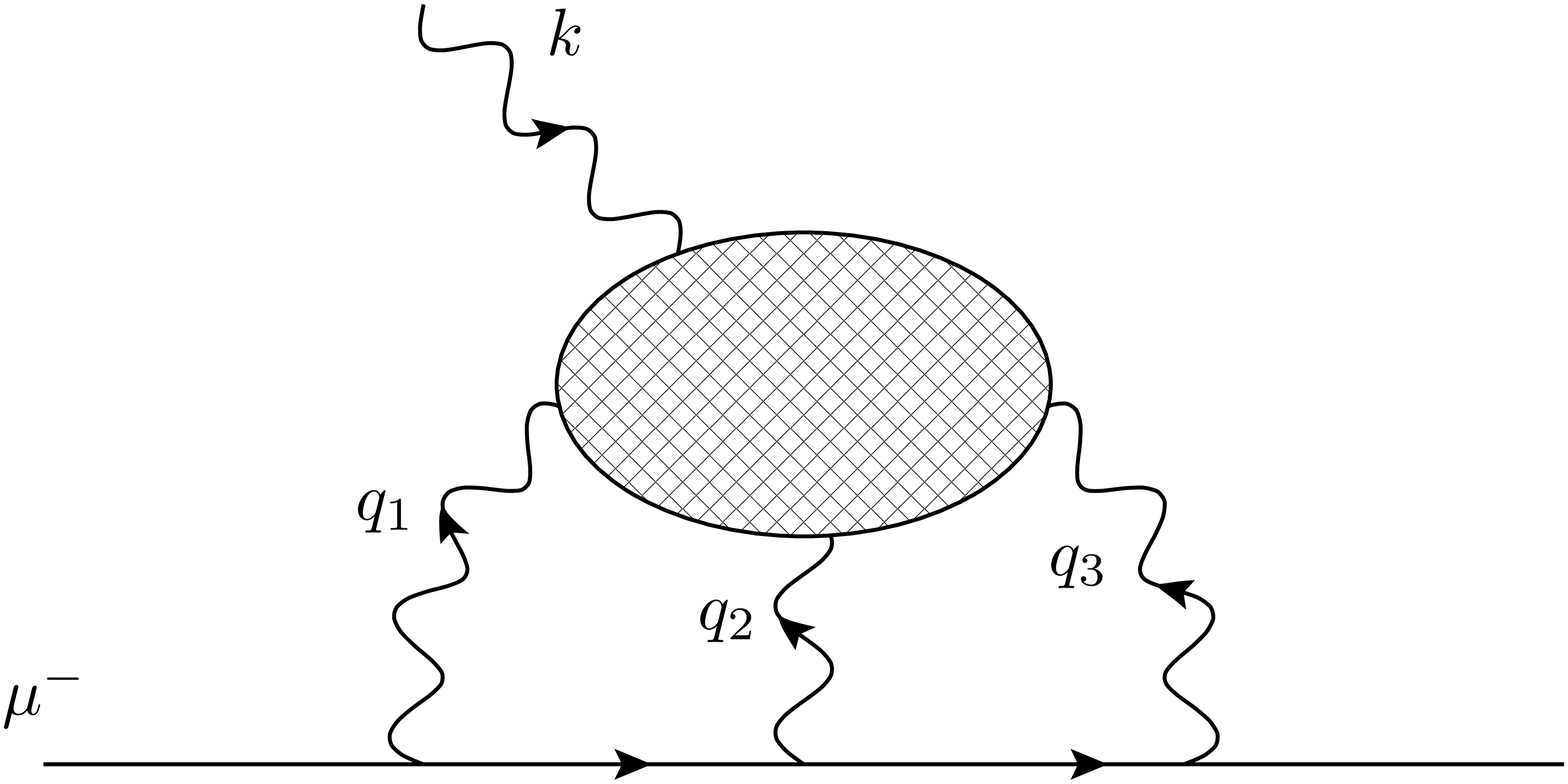}%
	\\
	\vskip 0.5cm
	\includegraphics[width=0.32\textwidth]{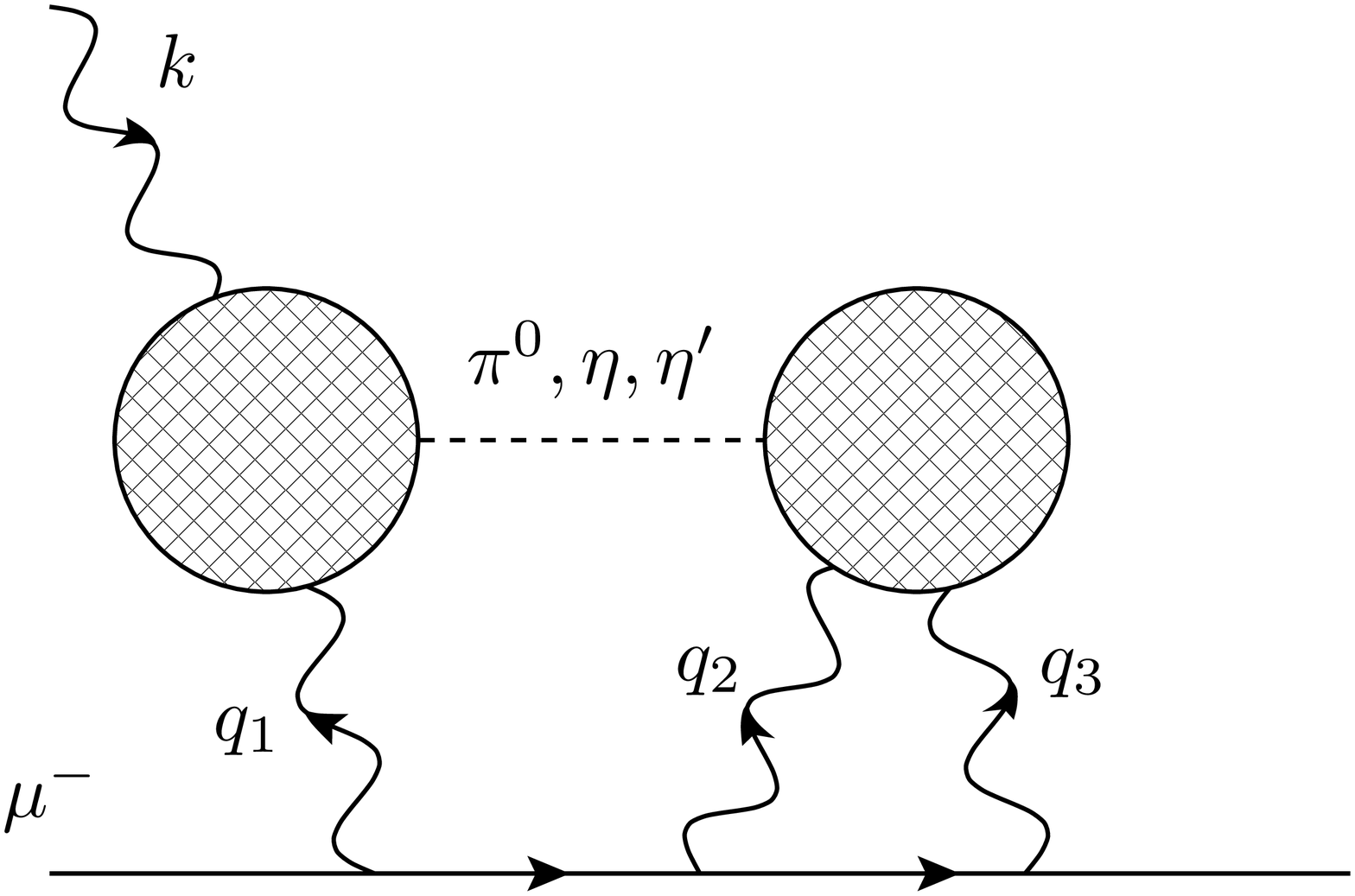}%
	\includegraphics[width=0.32\textwidth]{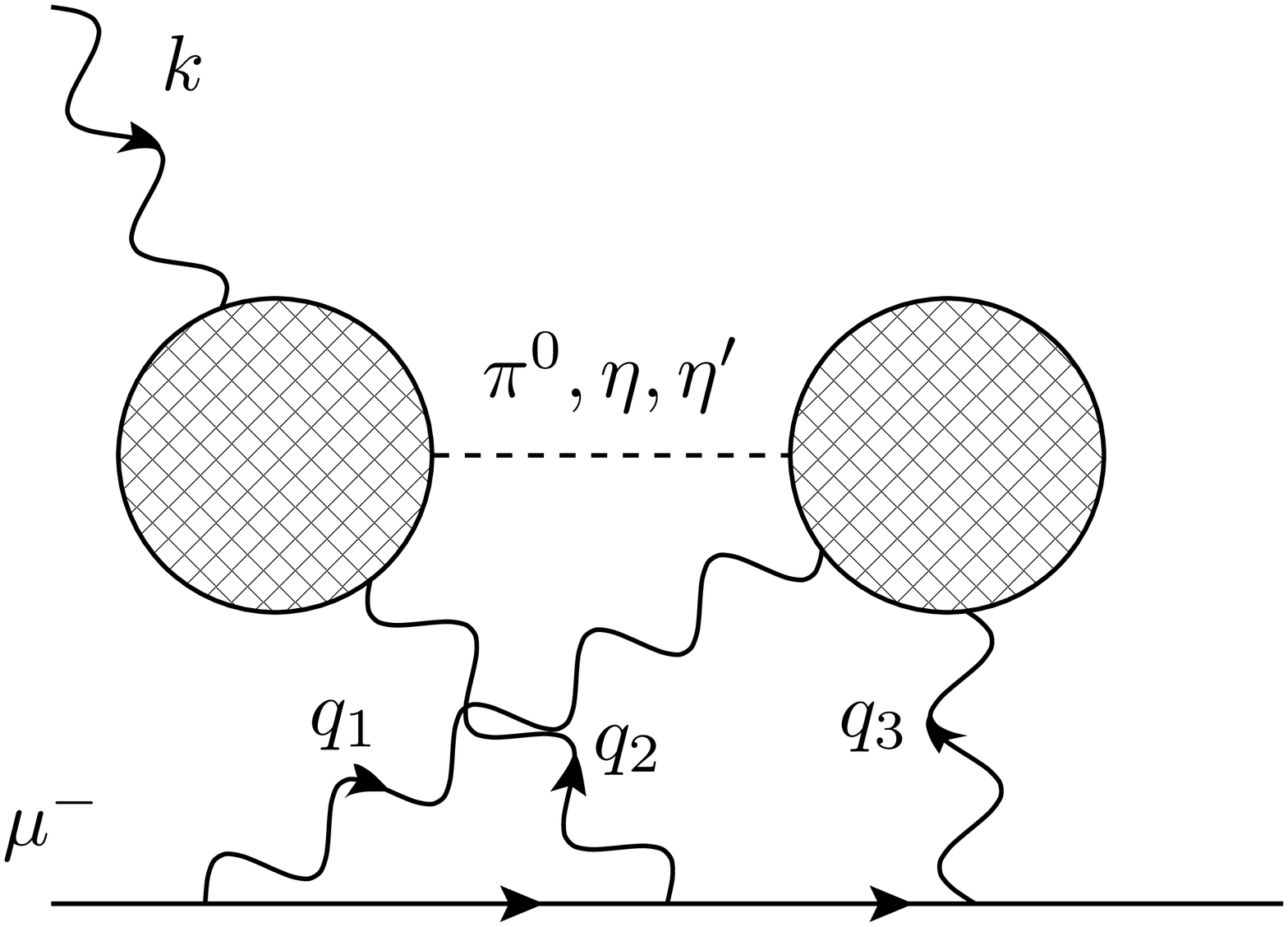}%
	\includegraphics[width=0.32\textwidth]{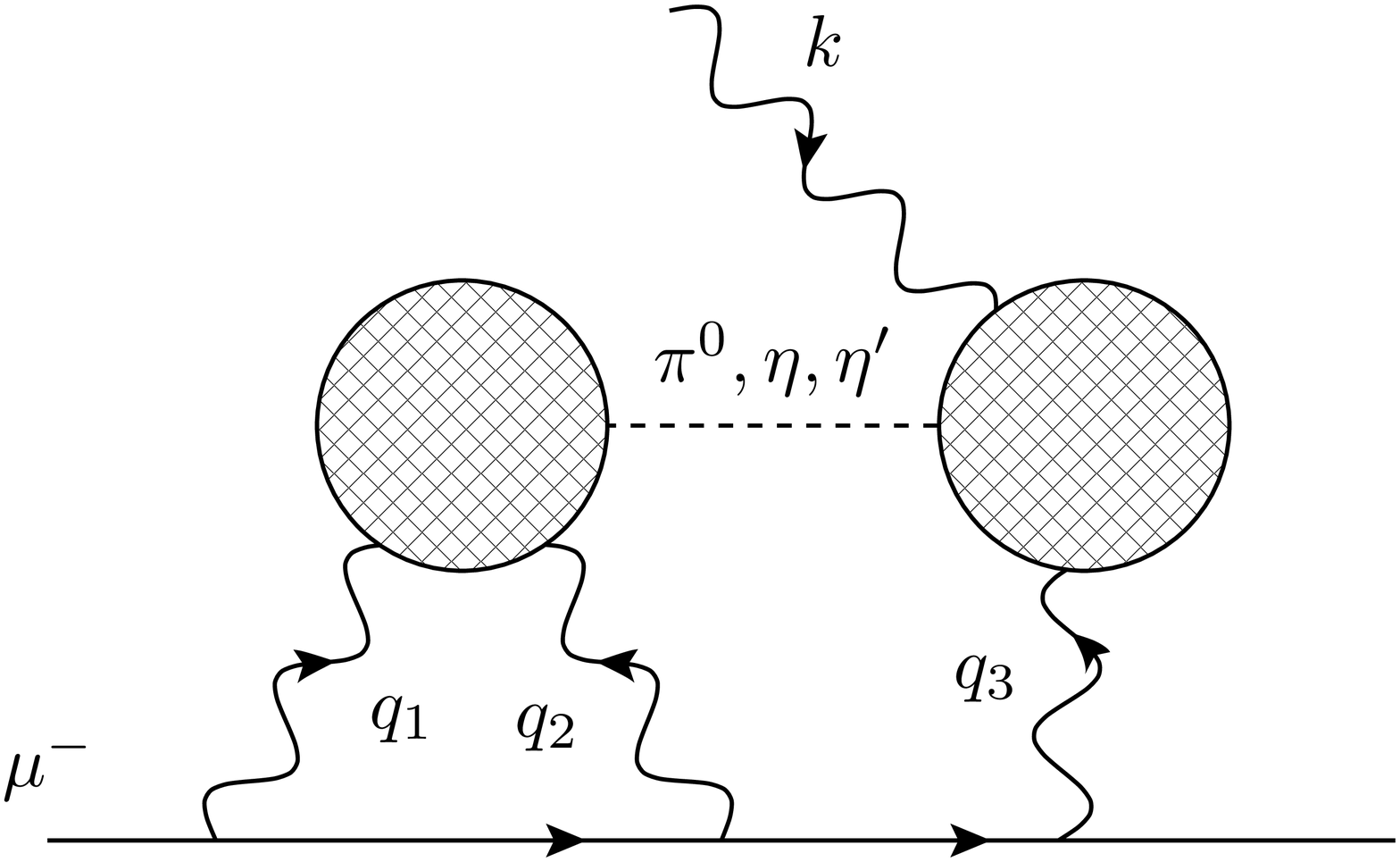}%
	\caption{\label{LBL diagram} Light-by-light correction (up) is supposed to be dominated by the diagrams, mediated by the pseudo scalar mesons (down).}
\end{figure}

While the electroweak corrections can be calculated very precisely~\cite{Aoyama:2008hz}, most uncertainties in the SM estimate of $(g-2)$ are coming from the hadronic corrections, which are essentially non-perturbative.  
The strong interaction contributions to the lepton magnetic moment consist of three pieces, 
the hadronic vacuum polarization,  the higher-order hadronic vacuum-polarization effect, 
and the hadronic light-by-light (LBL) scattering.  The contribution of the hadronic  vacuum polarization is the leading $\mathcal{O}(\alpha^2)$ correction of strong interactions to  $a_{\mu}$ and has been recently calculated in lattice~\cite{Aubin:2006xv}, but its present uncertainty is  about 10 times  larger than that of the current  experiment.  
Fortunately, however, one can bypass the calculation of the hadronic-leading-order (HLO) correction~\cite{Davier:2007ua} and use the experimental result of $e^+e^-\rightarrow\textrm{(hadrons)}$ (Fig.~\ref{HLO diagram}), which is related to the hadronic vacuum polarization by the unitarity and analyticity of the diagrams. The higher-order hadronic vacuum polarization effect can be calculated quite accurately, once the hadronic vacuum polarization is obtained. It is found to be $\delta a_{\mu}=-101(6)\times 10^{-11}$~\cite{Krause:1996rf}. 
Finally the hadronic light-by-light scattering  correction (Fig.~\ref{LBL diagram}) is the next-to-leading $\mathcal{O}(\alpha^3)$ effect,  but it is expected to be sizable to the current experimental accuracy, $6.3\times 10^{-10}$. It is therefore absolutely needed to estimate its effect accurately to assess the SM deviation of the muon $(g-2)$. 

There have been several attempts to estimate the hadronic light-by-light scattering corrections,  based on hadronic models or large $N_{c}$ approximations~\cite{Bijnens:2007pz,Jegerlehner:2009ry}. 
For last two decades  much improvement has been made in the LBL calculations despite a sign confusion~\cite{Hayakawa:2001bb}. Currently the positive sign is widely quoted for the pion-pole or  pion-exchange contributions to the hadronic LBL correction~\cite{Blokland:2001pb,Knecht:2001qg}, which is dominant in the large $N_c$ limit. Recently, however, a mistake related to the momentum conservation, which has not been considered seriously before, has been pointed out~\cite{Jegerlehner:2007xe} in the treatment of the $\pi^0\gamma^{*}\gamma^{*}$ form factor and is corrected properly~\cite{Nyffeler:2009tw}.

 The loop calculation of the LBL correction is  logarithmically divergent in a hadronic model, where a constant vertex is used instead of full three-point form factors. This implies that the LBL correction is sensitive to the choice of the regulator or the cutoff $\Lambda$ when we compare the model calculations with the data. In a model where the vector meson dominance (VMD) is adapted, however,  the mass of vector meson naturally regulates the ultra-violet (UV) divergences.   Among some models that regularize the hadronic uncertainty, the most popular ones are the meson dominance models like the VMD, the lowest meson dominance (LMD) and the LMD+V and so on, which fit among others the coefficient of QCD axial anomaly for the low energy on-shell photon~\cite{Hayakawa:1997rq,Knecht:2001qf,Melnikov:2003xd}.

In this work we study the hadronic LBL contributions in holographic models of QCD, which naturally incorporate the vector meson dominance. Holographic models  have been proposed recently for QCD~\cite{Sakai:2004cn,Erlich:2005qh}, inspired by the gauge/gravity duality, found in the string theory~\cite{Maldacena:1997re}. 

Several physical quantities of mesons and baryons such as their masses, couplings, and decay constants are calculated in holographic models of QCD and found to be in a good agreement with the experimental data~\cite{Sakai:2004cn,Erlich:2005qh,Hong:2006ta,Hong:2007kx}. 
Encouraged by the success of holographic QCD, we attempt to calculate  the hadronic LBL contributions to the muon ($g-2$). We consider in particular a hard-wall model of AdS/QCD,  defined in a slice of five dimensional  anti-de Sitter (AdS) spacetime, but our calculation can be easily applied to other holographic models of QCD.


\subsection{AdS/QCD and Light-by-Light scattering}
Solving QCD is very difficult, partly because at low energy  hadrons, rather than  quarks and gluons, are relevant degrees of freedom.  Holographic QCD is to describe QCD  directly with hadrons but in higher dimensions, prescribed by  the gauge/gravity duality. The extra dimension is related to the energy scale of QCD, the boundary theory. 
In the gauge/gravity duality, the global symmetries associated with conserved currents of the boundary gauge theory become the gauge symmetries in the bulk, while the gauge-invariant boundary operators are mapped to bulk fields.  

For a holographic model of QCD with three light flavors (up, down, and strange quarks) we consider 
a 5D $U(3)_L\times U(3)_R$ gauge theory in a slice of 5D AdS space-time, whose metric is given as, taking the AdS radius $R=1$,
\begin{equation}
ds^2=\frac{1}{z^2}(dx^\mu dx_\mu-dz^2),\quad \epsilon<z<z_{m}\,,
\end{equation}
where the ultraviolet (UV) regulator $\epsilon\to0$ and an infra-red (IR) brane is introduced at $z_{m}$ to implement the confinement of QCD~\cite{Erlich:2005qh}. The model, known as AdS/QCD,  is described by an action 
\begin{equation}
S=\int d^5x \sqrt{g}\textrm{ Tr}\left\{|DX|^2+3|X|^2-\frac{1}{4g_5^2}({F_L}^2+{F_R}^2)\right\}+S_{Y}+S_{CS}, 
\label{bulkaction}
\end{equation}
where $F$ is the field strength tensor of bulk gauge fields, $A$, dual of the QCD flavor currents, and the 5D gauge coupling, $g_{5}^{2}=12\pi^{2}/N_{c}$ ($N_{c}$ is the number of colors).
The bulk scalar $X$, which is bi-fundamental under the gauge group, is dual to the chiral-symmetry-breaking order parameter, $\bar q_{L}q_{R}$ at the boundary, and its 5D mass, $m^{2}=\Delta(\Delta-4)$, is related to the scaling dimension ($\Delta=3$) of the boundary dual operator. The covariant derivative is defined by $D_{M}X=\partial_{M}X-iA_{LM}X+iXA_{RM}$, where $A_{L(R)}$ denotes the  $\text{U}(3)_{L(R)}$ gauge fields.
To correctly reproduce QCD flavor anomalies, one has to introduce a Chern-Simons term~\cite{Domokos:2007kt},  given as
\begin{equation}
S_{CS}=\frac{N_{c}}{24\pi^{2}}\int\left[\omega_{5}(A_{L})-\omega_{5}(A_{R})\right]\,,\label{cs}
\end{equation}
where $d\,\omega_{5}(A)={\rm Tr}F^{3}$.  
Finally, for the anomalous ${\rm U}(1)_{A}$ we introduce  the bulk singlet $Y$, dual to $G_{\mu\nu}^{2}$ ($G\tilde G$) of gluon fields~\cite{Katz:2007tf}, described by
\begin{equation}
 S_{Y}=\int\,d^{5}x\sqrt{g}\left[\frac{1}{2}|DY|^2-\frac{\kappa}{2}(Y^{N_f}\textrm{det}(X)+\textrm{h.c.})\right]\label{5}\,,
\end{equation} 
where the singlet $Y$ has no mass term, since it is dual of dimension 4 operator, and $\kappa$ is a parameter to be fixed to give a correct mixing between $\eta$ and $\eta^{\prime}$.
The  bulk fields $A$, $X$ and  $Y$, that we introduced, are related to  towers of (axial) vector mesons~\cite{Erlich:2005qh} and (pseudo) scalar mesons~\cite{DaRold:2005vr}, respectively, upon Kaluza-Klein reduction. There are also higher dimensional operators like $F^{4}$ or $F^{2}|X|^{2}$ in the bulk action, but they are suppressed at low energy. (One could also introduce additional bulk fields to describe higher-spin states, but they are irrelevant to our discussion.)


According to the gauge/gravity duality, the classical bulk action becomes the generating functional for the one-particle irreducible (1PI) functions of the boundary gauge theory, once the source is identified as the bulk fields at the UV boundary.  Therefore the hadronic LBL diagram, which is  a four-point correlation function of the electromagnetic currents of quarks, 
can be easily calculated in holographic models of QCD, following the gauge/gravity correspondence. Since 
the quartic coupling for  the bulk vectorial $\text{U}(1)_{Q_{\textrm{em}}}$ gauge fields,  $V_{Q_{\textrm{em}}}$   ($Q_{\textrm{em}}=\frac{1}{2}{\bf 1}+I_{3}$), dual to the electromagnetic current of quarks,  are absent in the bulk action (\ref{bulkaction}),  
there are no 1PI 4-point correlators for the electromagnetic currents. Therefore in  AdS/QCD, where $F^{4}$ terms are suppressed, 
the LBL diagram  is just given  as a sum of 1PI three-point functions, 
connected by intermediate states~(FIG. \ref{LBL diagram}). The 1PI three-point functions consist of three different types; 
the vector-vector-scalar correlators,  the vector-vector-pseudo scalar correlators, and the vector-vector-axial vector  correlators. Among them the vector-vector-scalar correlators are suppressed since they come from the higher order terms like $F^{2}|X|^{2}$, while the vector-vector-pseudo scalar (axial vector) correlators are not suppressed as they
are given by  the bulk Chern-Simons (CS) term, Eq.~(\ref{cs}). 

Once the LBL diagram is calculated, it is straightforward to evaluate the LBL corrections to the muon 
($g-2$). Our calculation is similar to that of  Nyffeler~\cite{Nyffeler:2009tw} (see also~\cite{Knecht:2001qf}), where the pion contribution to the LBL correction was calculated most consistently.  The only difference is that the full off-shell anomalous form-factors for pseudo scalars are derived from AdS/QCD rather than constructed to satisfy the QCD constraints. As we will see later, our form-factor derived from AdS/QCD does satisfy the  asymptotic behavior for large and equal space-like photon momenta, obtained from perturbative QCD (pQCD), and also other asymptotic behavior derived from the operator-product expansion (OPE).

\subsection{$\pi^0$, $\eta$ and $\eta^\prime$ form factor calculation in AdS/QCD}
We consider the anomalous pion form-factor first. Since the pions are decoupled from the strangeness flavor, we need to consider  up and down flavors only, assuming equal mass.
The vacuum solution of the bulk scalar field $X$  is then written as  $\langle X \rangle=\frac{1}{2}(m_q z+\sigma z^3)\bf{1}_{2\times2}$, where $m_{q}$ and $\sigma$ correspond to the current quark mass and $\left<\bar q_{L}q_{R}\right>$, respectively. 
To analyze the correlation functions, we introduce  the vector and axial-vector gauge fields $V=(A_L+A_R)/2$ and $A=(A_L-A_R)/2$ and write $X= \left<X\right>\exp\left(2i\pi^{\hat a}t^{\hat a}\right)$, where $\hat a=S$ denotes the abelian part, while ${\hat a}=a\,(=1,2,3)$ are the ${\rm SU}(2)$ indices. (The generators $t^{S}={\bf{1}_{2\times2}}/2$ and $t^{a}=\sigma^{a}/2$, $\sigma^{a}$ denoting Pauli matrices). Since the abelian component $\pi^{S}$ is a part of $\eta$, it will be neglected for the discussion of the pion form-factor.   

In the axial gauge, $V_{5}=0=A_{5}$, 
the equations of motion for the transversal gauge fields are given as 
\begin{eqnarray}
	\left[\partial_z\left(\frac{1}{z}\partial_zV_\mu^{\hat a}(q,z)\right)+\frac{q^2}{z}V_\mu^{\hat a}(q,z)\right]_{\perp}&=&0\,,\label{1}\\
	\left[\partial_z\left(\frac{1}{z}\partial_zA_\mu^{\hat a}\right)+\frac{q^2}{z}A_\mu^{\hat a}-\frac{g_5^2v^2}{z^3}A_\mu^{\hat a}\right]_{\perp}&=&0\,,\label{2}
\end{eqnarray}
where $v(z)=m_q z+\sigma z^3$ and  $V_\mu^{\hat a}(q,z)=V_\mu^{\hat a}(q)V(q,z)$ and $A_\mu^{\hat a}(q,z)=A_\mu^{\hat a}(q)A(q,z)$ are the 4D Fourier-transform of vector and axial vector gauge fields, respectively.  The normalizable solutions to Eq.'s~(\ref{1}) and (\ref{2}) will correspond to the vector and axial vector mesons, respectively, satisfying the boundary conditions $V(q,\epsilon)=\partial_{z} V(q,z)|_{z_{m}}=0$ (same for the axial vectors). The boundary conditions for the source fields (the non-normalizable modes)  are on the other hand given as  $V(q,\epsilon)=1$ and $ \partial_zV(q,z)|_{z_m}=0$ and same for $A(q,z)$.
The pion fields come from both $\pi^{a}$ and the longitudinal components of the axial gauge fields, $A^{a}_{\mu\parallel}=\partial_{\mu}\phi^{a}$, which are related by equations of motion as, subjecting to  boundary conditions $\phi^{a}(q,\epsilon)=\partial_{z}\phi^{a}(q,z)|_{z_{m}}=0=\pi^{a}(q,\epsilon)$,   
\begin{eqnarray}
	\partial_z\left(\frac{1}{z}\partial_z\phi^a\right)+\frac{g_5^2v^2}{z^3}(\pi^a-\phi^a)&=&0\,,\label{3}\\
	-q^2\partial_z\phi^a+\frac{g_5^2v^2}{z^2}\partial_z\pi^a&=&0\,.\label{4}
\end{eqnarray}

\begin{figure}[tbh]
	\centering
	\includegraphics[width=0.5\textwidth]{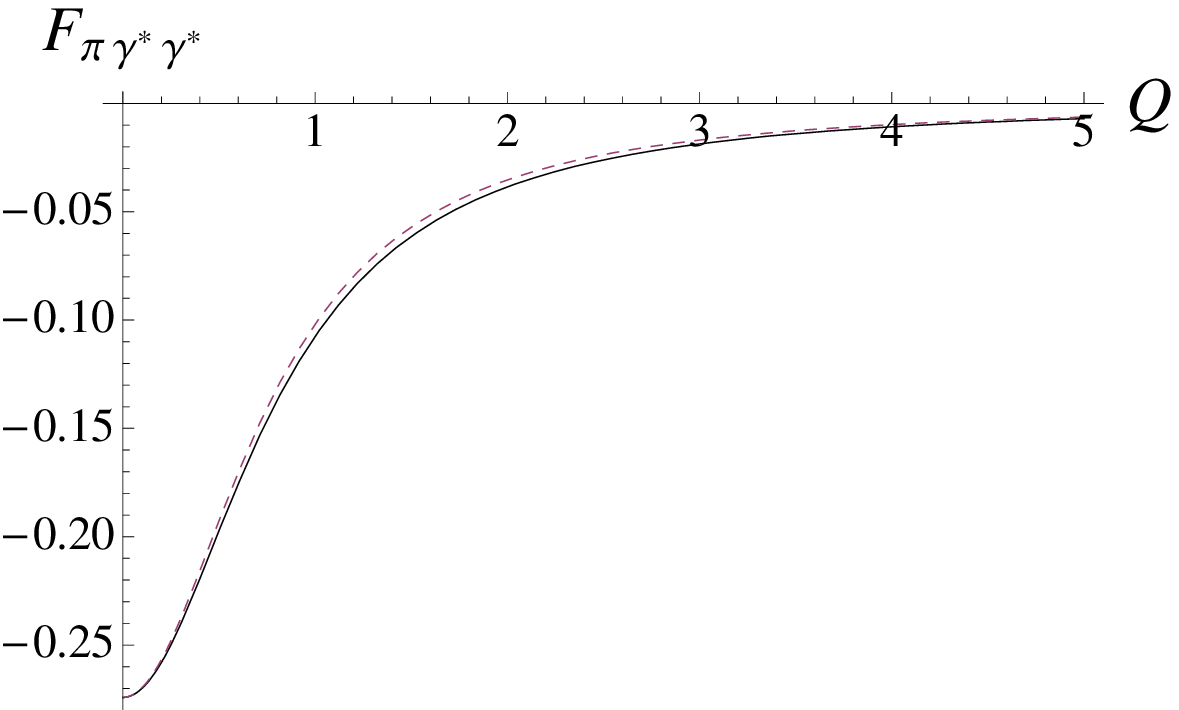}%
	\includegraphics[width=0.5\textwidth]{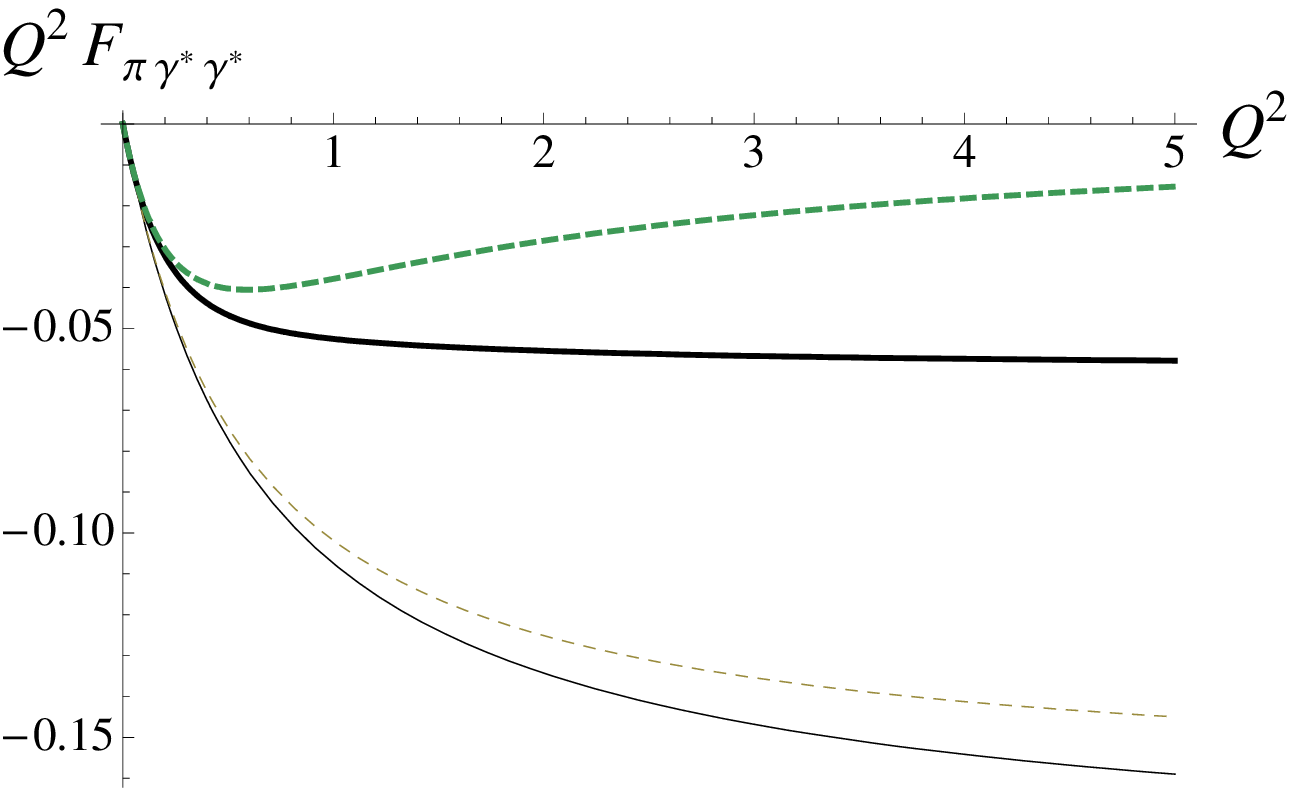}%
	\caption{\label{error estimation} The anomalous pion form-factors $F_{\pi\gamma^*\gamma^*}(Q,0)$ from AdS/QCD (solid) and VMD model (dashed) are presented in the left panel. The right panel displays both $F_{\pi\gamma^*\gamma^*}(Q,0)$ (lower two) and $F_{\pi\gamma^*\gamma^*}(Q,Q)$ (upper two) for both models. Notice that $F_{\pi\gamma^*\gamma^*}(Q,Q)$ grows as $\sim 1/Q^2$ asymptotically.}
\label{qcd}
\end{figure}

We can easily read off the anomalous pion form-factor   from the gravity dual,  by taking the functional derivation of the bulk Chern-Simons action, Eq.~(\ref{cs}) with the source field  at the UV boundary, $V_{Q_{\textrm{em}}}(q,\epsilon)$. For arbitrary external photon momenta $Q_{1}$ and $Q_{2}$, the pion momentum being $Q_{1}+Q_{2}$, the form factor is found to be 
\begin{equation}
	F_{\pi\gamma^*\gamma^*}(Q_1^2,Q_2^2)=\frac{N_c}{12\pi^2}\left[\psi(z_m)J(Q_1,z_m)J(Q_2,z_m)-\int dz \,\partial_{z}\psi(z)J(Q_1,z)J(Q_2,z)\right].\label{pionff}
\end{equation}
where $\psi(z)$, the wavefunction difference between $\phi^{a}$ and $\pi^{a}$,  is the wavefunction of non-normalizable pion mode with boundary conditions, 
$\psi(\epsilon)=1$ and $\partial_{z}\psi(z)|_{z_{m}}=0$ and $J(Q,z)$ is the Wick-rotated expression of the solution $V(q,z)$ defined below Eq.~(\ref{2}).
The anomalous pion form-factor is shown in Fig.~\ref{qcd} for various kinematic regions. 
As argued in~\cite{Grigoryan:2008up}, we find the pion form factor in AdS/QCD is in good agreement with the Brodsky-Lepage behavior, as shown in the  Fig.~\ref{qcd}, and also with the asymptotic behavior of pQCD for large space-like momenta of two photon legs, shown in the right panel of Fig.~\ref{qcd}. We also note that our pion form-factor gives  approximately the same magnetic susceptibility of quark condensate, obtained by OPE~\cite{Gorsky:2009ma}.

To calculate the anomalous form-factors for $\eta$, $\eta^{\prime}$, 
we need to consider the three-flavor case ($N_{f}=3$) and introduce  the term, needed for the $\text{U}(1)_{A}$ anomaly, included in the bulk action of flavor-singlet $Y$ field, $S_{Y}$, as  $\eta$ and $\eta^\prime$ have mixing. 
The bulk gauge fields are now generalized to $\text{U}(3)_{L}\times\text{U}(3)_{R}$ and 
$\langle X\rangle = \frac{1}{2}(Mz+\Sigma z^3)$ with
\begin{equation}
M=\left( \begin{array}{ccc}
m_q&0&0\\
0&m_q&0\\
0&0&m_s\end{array} \right)\quad\textrm{and}\quad\Sigma=\sigma\times\bf{1_{3\times3}}\,,
\end{equation}
satisfying the similar equations as Eq.'s~({\ref{1}) and (\ref{2}) except now $v(z)/2$ is replaced by the three-flavor vacuum solution $\left<X\right>$.

By solving the equations of motion for the pseudo scalars in the axial gauge,  which are the phase fluctuations of bulk scalars, $X$ and $Y$, and the longitudinal components of the axial gauge fields,   the wave functions of $\eta$ and $\eta^\prime$, that give the canonical kinetic terms for $\eta$ and $\eta^{\prime}$, can be found for an appropriate choice of $(m_\eta,\kappa)$ (FIG. \ref{Eta mass})~\cite{Katz:2007tf}.
\begin{figure}[htb]
	\includegraphics[width=0.8\textwidth]{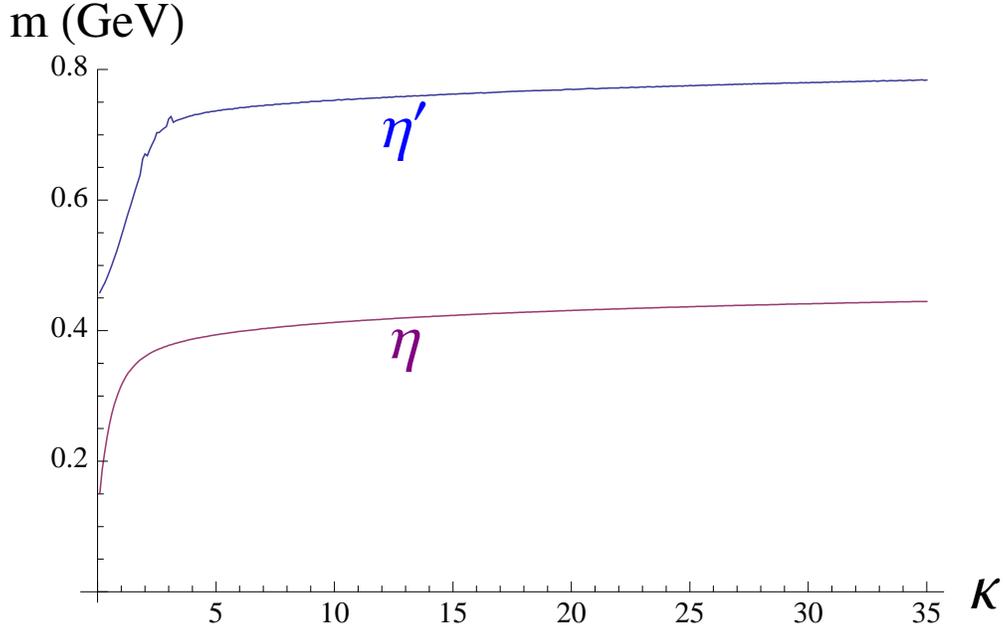}%
	\caption{\label{Eta mass} The set of $(m,\kappa)$ which gives the canonical four dimensional kinetic term of the $\eta$ and $\eta^\prime$ fields. Several masses are found for each $\kappa$ as a tower of the bound state of $I^G(J^{PC})=0^+(0^{-+})$, but only ground states are shown here.}
\end{figure}
Once the correct wavefunctions are found, not only the amplitude of $\eta, \eta^\prime\rightarrow\gamma\gamma$ processes, but the whole momentum structure of the form factors can be surveyed through the same method used above to deduce the anomalous pion form-factor. 
Namely, the form factors 
 are given similarly to Eq.~(\ref{pionff}) except that the  non-normalizable pion wavefunction $\psi(z)$ is replaced by  those non-normalizable wavefunctions of $\eta$, $\eta^{\prime}$ with different  overall coefficients. 
As shown in FIG.~\ref{form factor of eta and eta prime}, their profiles are quite similar to those of pions. (Note also the anomalous form factors are correctly normalized to be consistent with the QCD anomalies.) Combining all these, we can now calculate the 
pseudo scalar contributions to the LBL corrections for the muon  ($g-2$), following the recent work by Jegerlehner and Nyffeler~\cite{Jegerlehner:2009ry}.
\begin{figure}
	\includegraphics[width=0.8\textwidth]{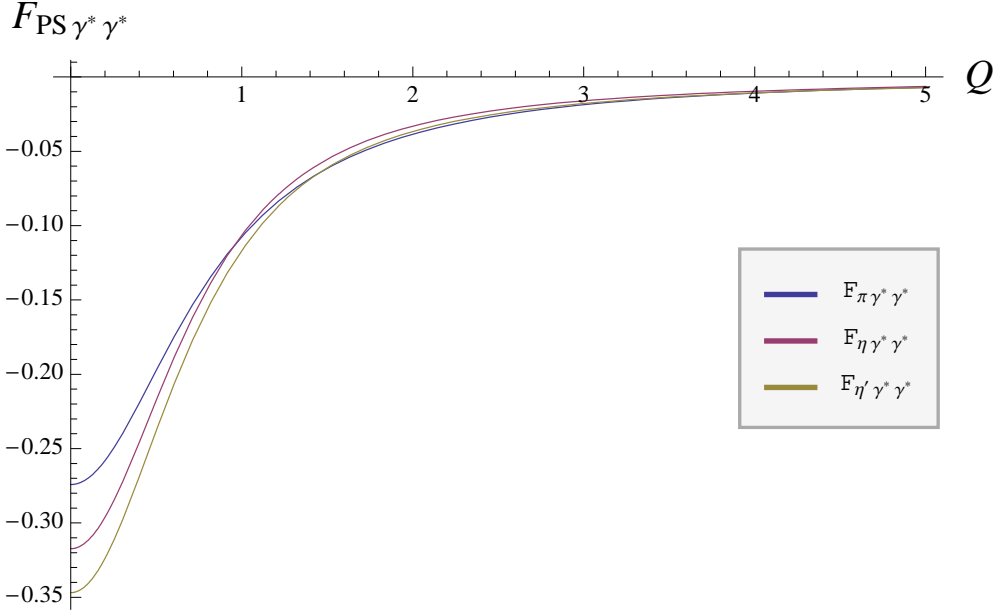}%
	\caption{\label{form factor of eta and eta prime} Pseudo scalar form factors $F_{\textrm{PS}\gamma^*\gamma^*}(Q,0)$ from the AdS/QCD are plotted. $Q$ is a Euclidean momentum. Note that for $Q=0$ their offset values are well matched to the experiments \cite{Donoghue:1992dd}.}
\end{figure}

\subsection{Results and Discussion}

Numerical calculations in Table \ref{final data} has been performed with \emph{Mathematica 6.0}, using the adaptive MonteCarlo scheme. To utilize the method used in Ref.~\cite{Jegerlehner:2009ry,Nyffeler:2009tw} for our calculation of meson exchange contributions, we have decomposed the source field of the vector gauge fields as follows~\cite{Hong:2004sa}:
\begin{equation}
	J(-iQ,z)=V(q,z)=\sum_{\rho}\frac{-g_5f_{\rho}\psi_\rho(z)}{q^2-m_\rho^2+i\epsilon}
	\label{modeexpansion}
\end{equation}
where $\psi_\rho(z)$ are the normalizable modes, corresponding to the (excited) rho mesons, with the boundary condition $\psi_\rho(\epsilon)=0$ and $\partial_z\psi_\rho(z)|_{z_m}=0$ and $f_\rho$ are their decay constants. 
Similarly we have expanded the  bulk axial gauge fields $A(q,z)$ for $\text{U}(3)$ in terms of normalizable modes, which contain  $\pi^0$, $\eta$, $\eta^{\prime}$ and $a_1 (1230)$, and towers of excited axial vector mesons.  
For our calculations we have set the parameters by $z_m=1/0.323$, $m_q=0.00222$, $m_s=0.04$, $\sigma=0.333^3$ and $\kappa=35$ where the energy scale is in GeV unit. 
We truncate for each photon-line the number of vector mesons in Eq.~(\ref{modeexpansion}) up to 4,  6, and 8 for our calculations. Beyond 8th modes, the form factor changes very little, as shown in Fig.~\ref{mode}. 
\begin{figure}[tbh]
	\centering
	\includegraphics[width=0.5\textwidth]{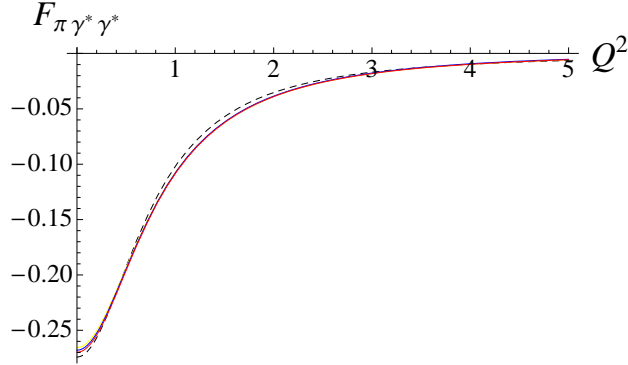}%
	\caption{\label{error estimation1}  Decomposing the source terms of vector gauge fields into the normalizable modes affects $F_{\pi\gamma^*\gamma^*}(Q,0)$ very little. The form-factor from VMD curve is dashed; yellow curve for 20 modes, blue for 60, and red for 150, respectively.}
	\label{mode}
\end{figure}
We have shown our results in Table~\ref{result}, which are close to the recently revised value, $a_\mu^{\textrm{PS}}=9.9\times10^{-10}$ in the LMD+V model by Nyffeler~\cite{Nyffeler:2009tw},  which correctly took into account of momentum conservation.
\begin{table}[tbh]
	\caption{\label{final data} Muon $g-2$ results from the AdS/QCD. ($10^{-10}$ factor should be multiplied to each number for $a_{\mu}$.) }
\begin{tabular}{ccccccccccccc}
\hline\hline
Vector modes & & $a_\mu^{\pi^0}$ & & $a_\mu^{\eta}$ & & $a_\mu^{\eta^\prime}$ & & & & $a_\mu^{\textrm{PS}}$\\
\hline
4 & & 7.5 & & 2.1 & & 1.0 & & & & 10.6\\
\hline
6 & & 7.1 & & 2.5 & & 0.9 & & & & 10.5\\
\hline
8 & & 6.9 & & 2.7 & & 1.1 & & & & 10.7\\
\hline\hline
\end{tabular}\label{result}
\end{table}

To conclude, we have calculated the pseudo scalar contributions to the LBL corrections to the muon ($g-2$) in an AdS/QCD model with three light flavors. Our holographic estimate gives results consistent with the recent estimate~\cite{Nyffeler:2009tw}, based  LMD+V model. Our approach  has a few parameters, which are highly constrained by low energy data of QCD, and is based on a principle, known as  gauge/gravity duality, valid  in the large $N_{c}$ and large 't Hooft coupling ($\lambda$) limit. 
Our result is therefore subject to $1/N_{c}$ and $1/\lambda$ corrections, which are at most $30\%$. 

The holographic models of QCD generically show that the LBL diagram is given by a sum of 1PI three-point functions, connected by intermediate states of pseudo scalars and axial vectors.
For the LBL corrections, we have considered light pseudo scalars ($\pi^{0}$, $\eta$, and $\eta^{\prime}$) only, since  the  axial vector meson, $a_1(1230)$, and its excited states are expected to be less important than light pseudo scalars. (A related work is in progress.) Finally we have calculated the full off-shell anomalous form-factors of light pseudo-scalars and found that they show the correct asymptotic behavior at large photon virtuality, given by pQCD.

\subsection{Acknowledgements}
We are  grateful to T. Kinoshita for useful comments and thankful to 
A. Nyffeler for very helpful communications. 
The work of D.K.H. was supported by the Korea Research Foundation Grant  funded by the Korean Government (KRF-2008-341-C00008) and also by grant  No.
R01-2006-000-10912-0 from the Basic Research Program of the Korea
Science \& Engineering Foundation. D. K. was supported by KRF-2008-313-C00162.
\subsubsection{}

\end{document}